\documentstyle[multicol,aps,prb,epsf]{revtex}

\def\n#1{{\bf #1}}
\begin{document}
\date{29 September 2000}

 \draft

\title{\bf Interaction potential between dynamic dipoles:
polarized excitons in strong magnetic fields}

\author{M. A. Olivares-Robles and S. E. Ulloa \\
 Department of Physics and Astronomy, and Condensed Matter
and Surface Sciences Program \\ Ohio University, Athens, Ohio
45701--2979}

\maketitle
\begin{abstract}

The interaction potential of a two-dimensional system of excitons
with spatially separated electron-hole layers is considered in
the strong magnetic field limit. The excitons are assumed to have
free dynamics in the $x$-$y$ plane, while being constrained or
`polarized' in the $z$ direction. The model simulates
semiconductor double layer systems under strong magnetic field
normal to the layers. The {\em residual} interaction between
excitons exhibits interesting features, arising from the coupling
of the center-of-mass and internal degrees of freedom of the
exciton in the magnetic field. This coupling induces a dynamical
dipole moment proportional to the center-of-mass magnetic moment
of the exciton. We show the explicit dependence of the
inter-exciton potential matrix elements, and discuss the
underlying physics.  The unusual features of the interaction
potential would be reflected in the collective response and
non-equilibrium properties of such system.

\end{abstract}



\section{Introduction}

The availability of new materials and interfaces in semiconductors
has allowed the exploration of novel electronic systems with
fascinating physical behavior.  Of particular relevance to the
model studied in this paper are the structures achieved by clever
use of multi-layer geometries, yielding double quantum wells, and
heterojunction interfaces of type II. In those systems, either by
the application of external electric fields or by the intrinsic
structure potentials, it is possible to achieve separation of
electrons and holes into distinct parallel layers, while
controlling the in-plane carrier densities.

This situation, of spatially separated electron and hole layers
has attracted the attention of several groups, both in theory and
experiment. The early proposals of Kogan and Tavger, \cite{kogan}
as well as Lozovik and Yudson, \cite{lozovik} and Shevchenko,
\cite{shevchenko} were focused on the possible correlations in
such systems due to the electron-hole interactions across the
layers.   More recently, other authors have theoretically explored
different features of these systems, from possible vortices,
\cite{vortices} and dark excitonic states due to hidden
symmetries, \cite{new-dzyubenko} to the various non-trivial
thermodynamic phases of these systems. \cite{Lozovik1999}  On the
experimental side, there has been substantial activity as well.
The experiments of Fukuzawa and coworkers gave tantalizing
evidence for the anticipated Bose condensation of `spatially
indirect' excitons in double quantum wells under strong electric
fields. \cite{Fukuzawa90}  Although later work has shown that the
interpretation of those results was not reliable, \cite{kash}
given the characteristics of the samples used, the concepts of
achieving Bose condensation of excitons in quantum wells is sound,
experimentally feasible, and currently being pursued in new
geometries and systems. \cite{snoke}

Controlled electron-hole separation in different layers/planes has
also been achieved using heterojunctions of type II, such as those
formed between InAs and AlSb (or GaSb).  In these structures, the
band alignments are such that electrons and holes are spatially
separated in equilibrium, as the bottom of the conduction band on
one side of the heterojunction lies lower than the top of the
valence band on the other. Butov {\em et al}. \cite{butov} have
reported photoluminescence experiments in AlAs/GaAs
heterojunctions, and their results suggest the appearance of a
Bose condensate in at least the high magnetic field regime.  Kono
{\em et al.} have also reported interesting spectroscopic data
suggesting an infrared-active state in the
InAs/Al$_x$Ga$_{1-x}$Sb system with unusual properties,
reminiscent also of those of a condensate. \cite{kono} Although
the carriers in these latter systems are not introduced optically
(as in the experiments above), the close proximity of carriers
{\em across} layers, while remaining at relatively low densities,
may yield exciton-like bound states of electrons and holes.

Depending on the details of each system, one can identify suitable
conditions under which the electron-hole layers would be well
described as a collection of polarized exciton-like dipoles.
\cite{lozovik} These conditions require the in-plane separation
of charge carriers to be much larger than that across the layers
(so that the electron-electron or hole-hole distances in each
plane are larger than those between electron and hole planes).
This in turn yields a system of excitonic {\em dipoles}
predominantly polarized along the normal to the interface. We
present here a study of the interactions between the resulting
exciton states, taking into account both the presence of an
intense magnetic field, and the internal structure of the
electron-hole pair.  The presence of the magnetic field
introduces a {\em dynamical} coupling between the center of mass
of the exciton and its relative coordinate, so that the
exciton-exciton scattering is a much more complicated event than
that occurring between point charges. We discuss here these
interactions, the potential characteristics and different
scattering events possible.

We should mention that perhaps the closest analog of these
polarized interacting excitons is that provided by polar
molecules, such as CO or HF.  A large number of theoretical and
experimental studies of the scattering events between such
molecules exists in the literature. \cite{HFscatt}  Although such
systems have permanent dipole moments and live in three
dimensions, the most different aspect to the excitons here is the
non-trivial coupling of their internal degrees of freedom with the
center-of-mass magnetic momentum.  This feature adds a very
interesting and subtle complication to the excitonic systems
studied here.

In fact, we will show that unlike more compact composite objects,
or in the polar-molecule analog, the scattering events here can
{\em strongly} affect the {\em internal} state of the
participating excitons.  In fact, as the in-plane dipole moment
of the exciton is proportional to its center-of-mass magnetic
momentum, the scattering will in general re-orient the dipole in
a well-defined way which depends on the momenta of the
participating excitons.  The event may also cause transitions to
{\em excited} internal states of the exciton, just as in the
polar molecule analog, although those may be suppressed here by
the strong magnetic field. This article studies the details of
such momentum exchanges and effective interactions, and provides
explicit expressions for the lowest matrix elements. Apart from
describing an unusual and interesting situation, these
interactions would play a vital role in a description of the
collective modes of this interacting boson gas.

It is also interesting to note the similarity of these exciton
dipoles with those believed to exist in the quantum Hall regime
at half-filling of Landau level.  The composite fermions there
develop a dipole moment proportional to the momentum, in a
similar way to the excitons we describe.  Although the underlying
physics is quite different, the scattering events of the
effective quasi-particles are possibly quite similar.
\cite{CFdipoles}  Perhaps some of the intuition developed in our
study of excitons would be of some use in better understanding
composite fermions in that regime.

In what follows, the specifics of the model are described in
section \ref{sect-model}, including a description of the role of
magnetic field in coupling the various degrees of freedom. Section
\ref{sect-potential} describes then the potential matrix elements
for the two-exciton scattering events.  Section \ref{cuatro}
illustrates the resulting scattering potential by considering a
few special events.  Section \ref{ultima} closes the work with
discussion and conclusions.

\section{Model} \label{sect-model}
\subsection{Exciton wavefunctions}

The system of interest can be characterized as a gas of electric
dipoles, which are free to move on the $x$-$y$ plane and are
effectively polarized either by the application of an electric
field in the $z$ direction, or by the built-in heterojunction
potentials of a type II system, as described in the introduction.
For concreteness and simplicity, we shall consider a model where
the electron and hole layers are separated by a set distance $d$,
and assume that their $z$-axis dynamics is strongly confined.
Consequently, the effective layer width of each of the layers is
assumed to be so narrow that the carriers have only
two-dimensional dynamics. This assumption, reinforced by the
electron-hole interaction, implies that the wave function spread
for both electrons and holes in the $z$-direction is negligible,
and that the other `transverse' states are so high in energy as
to be inaccessible for typical situations.  We assume further that
there is no electron tunneling into the hole layer and vice
versa. This is in fact the situation for type II heterojunctions
due to the built-in potentials, and also for indirect excitons in
double quantum well systems under strong {\em electric} fields.
(In the latter, however, the long-lived indirect excitons
co-exist initially with the short-lived direct-excitons created
during optical pumping. \cite{Fukuzawa90}) These physical
considerations can be suitably represented by constraining the
motion of electrons and holes to regions $z_{e}\leq 0$, and
$d\leq z_{h}$, respectively. This approximation neglects small
wave function penetration in realistic systems, but given typical
parameters, the penetration is small. \cite{YuCardona}

Although electron and hole cannot overlap in this simplified model
of the interface, they still interact via their Coulomb
attraction, and are able to form a system of spatially separated
but bound (if weakly) excitons. As mentioned above, this picture
of nearly-isolated and well-formed excitons should be an
appropriate description whenever the inter-particle distance in
the plane is much larger (low density) than the electron-hole
separation across the interface.

The wave function for each electron-hole pair in the system may
then be written as,
\begin{equation}
\psi({\bf r}_e, z_e; {\bf r}_h, z_h) =\Psi ({\bf r}_e,{\bf
r}_h)\delta (z_e)\delta (z_h-d), \label{wavefunction}
\end{equation}
where ${\bf r}_e$ and ${\bf r}_h$ are {\em two-dimensional
vectors} on the $x$-$y$ plane for the electron and hole,
respectively.  This factorization makes the problem effectively
two-dimensional.  Allowing for motion in the $z$-direction does
not alter qualitatively the two-dimensionality, but would require
the inclusion of a form factor to account for the finite
extension of the wavefunction in that direction.  This change
would only affect the results in a quantitative way, and can be
introduced straightforwardly, as done before in similar
situations.

The magnetic field is applied perpendicular to the layers, ${\bf
H}=H \hat{z}$, and it is assumed to be sufficiently strong that
the relation
\begin{equation}
r_{H}\ll a_{e}, a_h \, \label{limit}
\end{equation}
holds, where $a_{e,h}=\hbar^2\kappa/m_{e,h}e^2$ are the effective
Bohr radii for the electron and hole, $m_{e,h}$ are the effective
masses at $H=0$, $\kappa$ is the average background dielectric
constant, $e$ is the charge quantum, and $r_H = \sqrt{\hbar c/eH}$
is the magnetic length.  As discussed first by Gorkov and
Dzyaloshinskii, \cite{Gorkov} condition (\ref{limit}) allows one
to apply perturbation theory in this rather complex problem.
Motion of a 2D neutral electron-hole pair in a transverse magnetic
field ${\n H}=(0,0,H)$ is described by a Hamiltonian, which for
non-degenerate and isotropic bands in the effective mass
approximation reads
\begin{equation}
 H = \frac{1}{2m_e} (-i\hbar \nabla_e+\frac{e}{c}{\bf A}_e)^2
    +\frac{1}{2m_h} (-i\hbar \nabla_h-  \frac{e}{c}{\bf A}_h)^2-
    \frac{e^2}{\kappa \mid {\bf r}_e-{\bf r}_h\mid} \, .
\label{ec1}
\end{equation}
 Here, $\kappa=(\kappa _1+\kappa _2)/2$ is the average background
dielectric constant across the heterostructure.  Since $\kappa
_1\approx \kappa _2$ in typical systems, possible image charge
effects are small and neglected here.

The dynamics of the {\em single exciton} system is characterized
by a conserved quantity associated with the operator for magnetic
momentum of the center of mass, $\hat{\bf P}=-i\hbar\nabla_{\bf
R}-(e/c){\bf A}({\bf r})$. \cite{Gorkov}  Here,
\begin{equation}
{\bf R}= ( m_e {\bf r}_e+m_h {\bf r}_h)/M
\end{equation}
is the center of mass coordinate, while
\begin{equation}
{\bf r} = {\bf r}_e -{\bf r}_h
\end{equation}
is the relative coordinate of the electron-hole (e-h) pair,
$M=m_e+m_h$, and ${\bf A} ({\bf r})=\frac{1}{2}\n H\times {\bf r}$
in the Landau gauge.

The wavefunctions $\Psi _{nm\n P} $, describing the state of an
e-h pair in the field $\n H$ can be written as, \cite{Lerner}
\begin{eqnarray}
\Psi_{nm\n P}(\n r_e,\n r_h)&=&\exp\left\{\frac{i}{\hbar}\left(\n
P+\frac{e}{2c}\n H\times \n r \right)\cdot\n R\right\}\nonumber\\
&&\exp\left(\frac{i}{2\hbar}\gamma \n P\cdot \n r\right) \Phi_{nm}
(\n r-\n r_{\n P}), \label{2dwavefunction}
\end{eqnarray}
where $\gamma =(m_h-m_e)/M$, $\n P$ is the ``center-of-mass" or
``magnetic momentum" of the exciton associated with the operator
$\hat{\bf P}$,
\begin{equation}
 \n r_{\n P}= \frac{r_H^2}{\hbar} \, \hat{z} \times \n P \, ,
 \label{intrinsic}
\end{equation}
and the wave function $ \Phi_{nm}$ is identical to the
wavefunction of a charge $e$ in a field $\n H$, \cite{Landau}
\begin{eqnarray}
\Phi_{nm} ({\n r})&=&\left[\frac{n!}{2^{\mid m\mid+1}(n+\mid
m\mid)!\pi}\right]^{1/2}\frac{e^{-im\varphi}}{r_H}\times\nonumber\\
&& \left(\frac{\rho}{r_H}\right)^{\mid m\mid}L_n^{\mid
m\mid}\left(\frac{\rho^2}{2r_H^2}\right)\exp
{\left(-\frac{\rho^2}{4r_H^2}\right)} \, ,\label{movrel}
\end{eqnarray}
where $L_n^m$ are Laguerre polynomials, and $\rho=\mid \n r \mid$.
The wavefunctions $\Psi_{nm\n P}(\n r_e,\n r_h)$ describe then, in
the limit of high magnetic field, the dynamics of magnetoexcitons
with dispersion relation, \cite{Lerner}
\begin{equation}
\xi  _{nm}(P)=\xi _{nm}+E_{nm}(P) \, ,\label{spectrum}
\end{equation}
with
\begin{equation}
 \xi _{nm}=\hbar \omega _H \left( n+{1 \over 2} \right)
 (\mid m\mid-\gamma m+1) \, ,
\end{equation}
where the cyclotron frequency $ \omega _H=eH/\mu c$ is defined in
terms of the reduced mass of the exciton $\mu $, and
\begin{equation}
E_{nm}(P)= \langle \Phi _{nm} \mid -\frac{e^2}{\kappa \mid\n r +
\n r _{\n P} \mid}\mid \Phi _{nm} \rangle.
 \label{dispersion}
\end{equation}
Equation (\ref{spectrum}) is the energy of the $\Psi _{nm \n P}$
state to first order in the Coulomb interaction, and is a good
approximation as long as (\ref{limit}) holds. The states
constructed in this fashion can be viewed as an exciton that has
center-of-mass motion $\n P$ in the $x$-$y$ plane, and with
`internal structure' given by the state $\Phi _{nm} ({\n r} - {\n
r_{\n P}})$.

It is important to emphasize that the functions $\Phi _{nm}$ in
(\ref{2dwavefunction}) are centered at $\n r_{\n P}$, given by
Eq.\ (7), so that the actual in-plane separation between electron
and hole is $\langle \n r \rangle = \n r_{\n P}$, proportional to
its magnetic momentum. Notice moreover that the average radius
vector between electron and hole is orthogonal to $\n P$, {\em
for all} $n$ an $m$, and it vanishes for $P=0$. This peculiar
dependence of the dipole moment on the magnetic momentum can be
intuitively understood as the result of the Lorentz force tending
to separate the charges in each pair. Notice that the in-plane
polarization reduces their Coulomb interaction, and makes the
exciton more susceptible to ionization by system imperfections.
\cite{Lerner} This $P$-dependence will have also important
consequences for the scattering matrix elements, as we will see
later: even an {\em elastic} re-orientation of $\n P$ results in
a re-alignment of the dipole moment, which in turns changes the
interaction with other dipoles. Nevertheless, notice that because
this system has spatially separated electron-hole layers, the
{\em total} dipole moment vector of the exciton has a constant
component along the $z$-axis, and this is the dominant component
in most cases.

\subsection{Inter-exciton residual potential}

As already mentioned, the charge separation imposed by the
layered geometry produces an effective polarization, nearly
perpendicular to the interface for small $\n P$ values, and
resulting in a non-zero dipole moment for all the excitons
described here. This fact gives rise to an overall repulsive
interaction between all excitons in the system.  It will be this
``residual potential" that provides for a collective response, as
we describe in the next section.

To the lowest order in a multipole expansion, the residual
interaction potential between two excitons located at $\n x$ and
$\n x'$, respectively, can be written as the interaction between
two dipoles,
\begin{eqnarray} \label{eq:pot}
V& =& \frac{\n p \cdot \n p'}{\kappa \, \mid \n x - \n x' \mid^3
} -\frac{3 \,[\n p \cdot (\n x-\n x')][\n p' \cdot (\n x-\n
x')]}{\kappa \, \mid \n x - \n x' \mid^5 } \, .
\end{eqnarray}
The dipole moments $\n p$ are generated by the non-zero
expectation value of the relative coordinate. Correspondingly,
${\bf p}= e({\bf r} + \hat{z} d)$, where ${\bf r}$ is the
in-plane relative coordinate, and $d$ is the $z$-axis separation.
As this expression depends on both the relative and the
center-of-mass coordinates ($\n x$ and $\n x'$) it needs to be
evaluated for each exciton state wavefunction (see next section).

Notice that this dipolar approximation should be valid as long as
the exciton separation is larger than any of the characteristic
size-length scales of the excitons themselves,
 \begin{equation} \label{condition}
 |\n x  - \n x'| \gg |\langle {\bf r} \rangle |, d, a_\mu \, ,
 \end{equation}
 where $\langle {\bf r} \rangle$ is the in-plane exciton mean
radius ($={\bf r}_{\n P}$), $a_\mu = \hbar^2 \kappa / \mu e^2$ is
the exciton's Bohr radius, and $d$ is the $z$-axis e-h separation.
For closer inter-exciton separations, one should in principle
include higher multipoles in the interaction between excitons,
accounting for the constituent electrons and holes. The dipolar
approximation would break down as the in-plane carrier density
increases, violating the condition (\ref{condition}).
Correspondingly, this condition would require $| \n x - \n x' |
\gtrsim a_\mu \approx100$ {\AA} in typical materials/systems. The
in-plane densities would need to satisfy $n \ll 1/\pi a_\mu ^2
\approx 10^{12}$ cm$^{-2}$, quite a reasonable request (given the
typical experimental densities of $10^{10}$ to $10^{11}$
cm$^{-2}$). \cite{Fukuzawa90,kash,snoke,butov,kono}

\section{Interaction matrix elements}
\label{sect-potential}

\begin{figure}[b]
\centerline{\epsfxsize=10cm \epsfbox{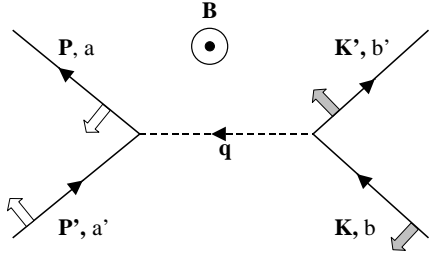}}
 \vspace{-2in}
\caption{A sketch of a scattering event between exciton dipoles.
One of the exciton-particles makes the transition $\{\n P',a'\}
\rightarrow \{\n P, a\}$, while the other one changes $\{\n K, b
\} \rightarrow \{ \n K', b'\}$. Notice that {\em in-plane} dipole
moments are perpendicular to each momentum, and they change upon
scattering (as indicated by blank and shaded arrows). Dashed line
indicates the momentum exchange due to the inter-exciton
potential.}
 \label{diagram}
\end{figure}

As discussed earlier, the main motivation for considering this
problem is to model the dielectric response function for a 2D Bose
gas of dipole-like polarizable bosons in a strong magnetic field.
In this context, the dielectric function in the self-consistent
field mean-field approximation can be written as,
\cite{SCFA,Dmitrii}
\begin{equation}
 \epsilon_{\alpha'\alpha ,\beta\beta'}(\omega) =
\delta_{\alpha' \beta'} \delta_{\beta \alpha} - V_{\alpha' \beta;
\alpha \beta'} \Pi_{\beta \beta'} (\omega) \, ,
\end{equation}
 where $\Pi_{\beta \beta'}$ is the polarization matrix, $\omega$ is
the frequency of the perturbing potential, and the inter-exciton
interaction potential matrix elements in the excitonic wave
function basis are given by
\begin{eqnarray}
V_{\alpha '\beta;\alpha \beta '}&=&\int
\psi ^*_{\alpha '}(\n {\Gamma} )\psi ^*_{\beta}(\n {\Gamma }')
V(\n {\Gamma },\n {\Gamma }')
\psi _{\alpha }(\n {\Gamma} ) \psi
_{\beta'}(\n {\Gamma }') d\n {\Gamma}  d\n {\Gamma '}.
\label{generalpotential}
\nonumber\\&&
\end{eqnarray}
Here, $\n {\Gamma}$ and $\n {\Gamma }'$ refer to the exciton
degrees of freedom, with $\n {\Gamma}= \{ \n r, \n R, z_e, z_h\}$, or $\{
\n r_e , \n r_h, z_e, z_h\}$, and the $\alpha$ and $\beta$ indices denote
the $\{nm \n P\}$ set of excitonic quantum labels.  Using the
states described above, we have for the first term of the
potential ($12$),
\begin{eqnarray}\label{term1}
V^{(1)}_{\alpha '\beta;\alpha \beta '}&=& \Big[{\n p}_{\alpha
'\alpha}\cdot {\n p}_{\beta \beta '} +e^2d^2M_{\alpha
'\alpha}M_{\beta \beta '}\Big]
\phi(q)\,\delta ({\n q}-{\n q'})
,\nonumber\\
\end{eqnarray}
while for the second term,
\begin{eqnarray}
V^{(2)}_{\alpha '\beta;\alpha \beta '}&=&\Big[({\n q}\cdot {\n
p}_{\alpha '\alpha})({\n q}\cdot {\n p}_{\beta \beta '})\chi
_1(q)\nonumber\\
&&-({\n p}_{\alpha '\alpha}\cdot {\n p}_{\beta \beta '})\chi
_2(q)\Big]\,\delta ({\n q}-{\n q'}) \label{term2} \, .
\end{eqnarray}
The delta functions in these equations ensure overall magnetic
momentum conservation in the scattering between two excitons,
i.e.,
\begin{equation}
\n P'+\n K=\n P+\n K' \, ,
\end{equation}
 with $ \hbar \n q = \n P - \n P'$, and $\hbar \n q' = \n K - \n
K'$, as one expects for translational invariant systems.  The
labels for incoming and outgoing momenta and other exciton quantum
numbers are shown in Fig.\ \ref{diagram}, where $\n q$ represents
the in-plane momentum transfer due to the scattering event
between excitons.  In this notation, the scattering process is
fully described by the change of the remaining internal state
labels, $a' \rightarrow a$, and $b \rightarrow b'$, as indicated
in the figure.

In these expressions, the functions $\phi(q)$, $\chi_1(q)$, and
$\chi_2(q)$, are Fourier transforms of the dipolar interaction
dependence on the center-of-mass coordinates, and are explicitly
written in Appendix A.  The in-plane ``dipole moment" matrix
elements are given formally by,
\begin{equation}
{\n p}_{\alpha '\alpha}=\frac{2e}{i\gamma}\frac{\partial
}{\partial \n q}M_{\alpha '\alpha} \, ,  \label{19}
\end{equation}
where $\hbar \n q = \n P - \n P'$ as above, and the non-local
``overlap" matrix elements are given by
\begin{eqnarray}
M_{{\alpha '}{\alpha}}&=&M_{n'm'\n P',nm\n P}\nonumber\\
&=&e^{i\frac{\gamma}{2}\n
q\cdot \n r_{\n P'}}\int {e^{i\frac{\gamma}{2}\n q\cdot \n r} \,
\Phi _{n'm'}^*(\n r) \, \Phi _{nm}(\n r-\n u)} \, d^2 \n r \, ,
\label{20}
\end{eqnarray}
with similar expressions for $ \n p _{\beta \beta'}$ and $M_{\beta
\beta'}$.  In the last equation, we have used $\n u =\n r_{\n P}
-\n r_{\n P'}$, and the gradient in Eq.\ (\ref{19}) refers to the
{\em explicit} $q$-dependence shown in (\ref{20}), different from
$\n u$.

In the following sections, we describe typical features of the
overlap and dipole matrices, $M_{\alpha \alpha'}$ and $\n p
_{\alpha \alpha'}$.

\subsection{Dipole matrix elements}

In order to better understand the nature of the residual
interaction matrices, we evaluate some of the lowest elements. In
what follows, and for notational convenience, we use $\alpha =
\{nm \n P \} = \{a \n P\}$, with $a$ standing for the indices of
the $\Phi_{nm}$ states, so that one can write for example,
\begin{eqnarray}
 \n p _ {\alpha' \alpha} &=&  \n p _{a' \n P', a \n P}
 = \n p _{a'a} (\n P - \n P',\n P + \n P' )
 = \n p _{a'a} (\n q,
 \n P') \, ,
\end{eqnarray}
where the explicit dependence on the sum and difference of
participating momenta is indicated.  The last equality uses the
overall conservation of magnetic momentum provided by the delta
functions in (\ref{term1}) and (\ref{term2}).

The simplest dipole moment matrix element (for $a=\{00\}=a'$) can
be written as (see Appendix B),
\begin{eqnarray} \label{p0000}
{\n p}_{00 \n P',00 \n P} = \n p _{00,00} (\n q,\n P') &=&
\frac{e}{2}\left (\n r_{2\n P'+\hbar \n q}+i\gamma r_H^2\n q\right
)\, e^{i\frac{\gamma}{4}\n q\cdot \n r_{2\n P'+\hbar \n q}} \,
e^{-(1+\gamma^2) q^2 r_H^2 /8} \, ,
\end{eqnarray}
 where the in-plane momentum exchange $\n q$, and the {\em
incoming} magnetic momentum $\n P'$ are used to specify the dipole
and non-local overlap matrix elements.

From this expression, the (non-interacting) long-wavelength limit
$\n q=0$ yields,
\begin{eqnarray}  \label{p00}
{\n p}_{00,00}(\n q=0, \n P')&=&\frac{e}{2} \n r_{2\n P'}=
\frac{er^2_H}{\hbar} \left[\hat{z} \times \n P' \right]\nonumber\\
& = &e \n r_{\n P'} \, .
\end{eqnarray}
This represents what one could call the `proper' dipole moment of
the exciton in state $nm=00$, and with magnetic momentum $\n P'$,
since the expectation value of the relative coordinate is $\n
r_{\n P'}$. In fact, it is possible to show (see Appendix B) that
{\em all} the diagonal dipole matrix elements in the limit $\n q
\rightarrow 0$ yield $\n p _{aa} (\n q=0, \n P) = e \n r _{\n P}$,
since in fact all $\Phi_{nm}$ states have the same dipole moment
(in this high field limit), as we discussed following Eq.\
(\ref{dispersion}). We emphasize that large momentum values
correspond to larger exciton size and lower binding energy, as
the exciton is increasingly polarized. \cite{Lerner}  One expects
that such high-$\n P$ states would be easily affected (even
disintegrated) by perturbations in the system, such as impurities
and surface inhomogeneities.

It is interesting to note the role that $\gamma = (m_h - m_e)/M$
plays in Eq.\ (\ref{p0000}), providing an imaginary part (or
phase) proportional to $\n q$ to the dipole matrix element. Notice
further that for any $\gamma$ values, a non-vanishing momentum
exchange $\n q$ depresses exponentially the dipole matrix element,
with a characteristic length $\approx r_H$. Since non-vanishing
$q$ corresponds to the momentum/energy transfer from one exciton
to the other, high momentum transfer processes will appear to be
strongly suppressed by this potential.  Let us discuss these
features in the next section.

\subsection{Potential matrix elements}

The simplest elements of the potential are those diagonal in the
$\{a,a'\}$ indices.  For two excitons with incoming momenta $\n
P'$ and $\n K$ which exchange momentum $\n q$, the potential
matrix element is given by,
\begin{eqnarray}
V_{00,00;00,00}(\n q,\n P',\n K)&=&\left\{
\phi(q)\left[\frac{e^2}{4}(\n r_{2\n P'+\hbar \n q}+i\gamma
r^2_H\n q)\cdot(\n r_{2\n K-\hbar \n q}-i\gamma r^2_H\n
q)+e^2d^2\right]+ {\mathcal V}_{00,00;00,00}(\n
q,\n P',\n K) \right\}\times\nonumber\\
&& \tilde{M}_{00,00}(-\n q,\n P')\tilde{M}_{00,00}(\n q,\n K)
\label{final2}
\end{eqnarray}
where,
\begin{equation}
\tilde{M}_ {00,00}(\n q,\n P')= \exp\left(i\frac{\gamma}{4}\n
q\cdot\n r_{2\n P'+\hbar \n
q}-(\gamma^2+1)\frac{r^2_Hq^2}{8}\right) \, , \label{M00P'}
\end{equation}
and,
\begin{eqnarray}
{\mathcal V}_{00,00;00,00}(\n q,\n P',\n
K)=\frac{e^2}{4}&\Big\{&\Big[\n q\cdot\Big(\n r_{2\n P'+\hbar \n
q}+i\gamma r^2_H\n q\Big)\Big]\Big[\n q\cdot \Big(\n
r_{2\n K-\hbar \n q}-i\gamma r^2_H\n q\Big)\Big]\chi_1(q)\nonumber\\
&&-\Big(\n r_{2\n P'+\hbar \n q}+i\gamma r^2_H\n q\Big)\cdot
\Big(\n r_{2\n K-\hbar \n q}-i\gamma r^2_H\n
q\Big)\chi_2(q)\Big\} \, . \label{v200}
\end{eqnarray}
Notice that these expressions contain a contribution from the
constant $z$-component of the dipole moment, $ed$, as well as from
the in-plane components.

From these equations, and considering the $\n q \rightarrow 0$
limit of the potential (see Appendix A), we may write
 \begin{equation}
V_{00,00;00,00}(\n q\to 0,\n P',\n K)= \frac{e^2}{2 \sqrt {\pi }d
\kappa} \left \{ \left( 1-6\,{\pi }^{3/2}\,\Gamma (3/4) \right) \,
\n r_{\n P'}\cdot\n r_{\n K}+ d^2 \right \} \, .
 \label{limitq0}
\end{equation}
This result is expressed in terms of the proper dipole moment of
each exciton, proportional to $\n r_{\n P'}$ and $\n r_{\n K}$,
as intuitively expected by Lerner and Lozovik. \cite{Lerner} It
is clear that the sign of the first term in the interaction
(\ref{limitq0}) depends on the relative orientations of $\n P'$
and $\n K$, and the resulting dipole moments. The total
interaction between excitons will be more/less repulsive for
anti/parallel $\n P'$ and $\n K$, as the contribution to the
$z$-axis moment is modulated by the in-plane component. For small
and moderate magnetic moment values, typical of the excitons in
this system at low temperatures, the repulsive interaction is
however only weakly modulated, since $d \gtrsim r_{2 \n P} =
r^2_H P $, but it is still dependent on the relative orientation
of the proper dipoles.

\section{Scattering events}
\label{cuatro}

As we have mentioned, the potential matrix elements above allow
the description of the collective excitations of the weakly
repulsive gas of polarized excitons, in a manner similar to that
treated in Ref.\ [\onlinecite{Dmitrii,Raul}].  Moreover, these
potential expressions can also be used in a quantitative
description of the kinematics of scattering events, as those
needed in a treatment of the distribution function via the
Boltzmann equation to evaluate drag,\cite{Boltzeq} or the
evaluation of the Bose condensate properties in this
dipole-interacting system.\cite{Goral-Santos} As an illustration
of their use, we describe in this section how the potential matrix
elements calculated above can provide rates and cross sections
for different inter-exciton scattering events.  For simplicity,
we deal here with `elastic' collisions, when there is no change of
the internal state under scattering.  More complex events are in
principle also allowed, although `inelastic' processes are
suppressed by the strong field.

If one considers scattering events in which the internal state of
the excitons is left unchanged, one is then faced with a purely
{\em kinematically elastic} collision.  The description of this
elastic collision, like any problem of two bodies, is simplified
by changing to a system of coordinates in which the center of
mass of the two particles is at rest. The scattering angle in the
center of mass reference frame is denoted by $\theta$, and it is
related to the angles $\theta_1$ and $\theta_2$ giving the
scattering angles of the two particles in the {\em laboratory}
system of coordinates. In the case in which the second particle
was at rest before the collision, for example, one can write
\cite{Landau}
\begin{displaymath}
\tan \theta_1=\frac{m_2 \sin \theta}{m_1+m_2\cos\theta} \, ,
\quad\quad \theta_2=\frac{1}{2}(\pi-\theta) \, ,
\end{displaymath}
where $m_1$, $m_2$ are the masses of the scattering objects. In
our case, the masses of the two `particles' (excitons) are the
same ($m_1=m_2=M^{exc}_{a}$, which depends on the internal state
$a=\{nm\}$), and we have simply,
\begin{displaymath}
\theta_1=\frac{1}{2}\theta \, , \quad\quad
\theta_2=\frac{1}{2}(\pi-\theta) \, ;
\end{displaymath}
so that the particles diverge at right angles in the laboratory
frame.

The scattering cross-section can be calculated using the Born
approximation, since the residual potential (\ref{eq:pot}), may
be considered a weak perturbation. Notice, furthermore, that the
residual potential depends only on the distance between excitons
$\n x-\n x'$, so that the scattering field is central. Now, in
the center of mass frame of reference, we can write
\begin{equation}
\n P'_{CM}=\n n \qquad\qquad \n K_{CM}=- \n n \, ,
\end{equation}
where $\n n=(\n P'-\n K)/2$, is the relative magnetic momentum
between excitons. Thus, the interaction potential matrix element
for this event, where the internal state is assumed to be
$a=\{00\}$ before and after the collision, can be written as
\begin{eqnarray}
V_{00,00;00,00}^{CM}(\n q,\n n)&=& \frac{e^2}{4} \left\{
4\phi(q)d^2- \Big[\chi_2(q)-\phi(q)\Big]\Big(\n r_{2\n n+\hbar \n
q}+i\gamma r^2_H\n q\Big)^2- \Big(\n q\cdot \n r_{2\n n}+i\gamma
r^2_H\n q^2\Big)^2\chi_1(q)\right\} \times \nonumber \\
& & e^{-(\gamma^2+1)\frac{r^2_Hq^2}{4}} \, ,
 \label{final3}
\end{eqnarray}

\noindent where $\hbar \n q=\n P-\n P'=\n K - \n K'$.  Other
internal states are given by a different detailed expression, but
identical kinematics (Appendix B).

In the case under consideration, Eq.\ (\ref{final3}) describes the
matrix element for a `transition' (scattering event) from a state
with momentum $\n n$ to the state with momentum $\n n'=(\n P -\n
K')/2= \n n + \hbar \n q$, which we could then denote as
$U_{nn'}$. Correspondingly, the scattering rate can be calculated
from the golden rule,
\begin{equation}\label{rate}
dW_{fi}=(2\pi /\hbar)|U_{nn'}|^2\delta (E_n'-E_n) \, ,
 \label{taza}
\end{equation}
where the final and initial energies of the exciton of interest
are,
\begin{equation}
E_n'-E_n=(\n n'^2-\n n^2)/2M^{exc}_{00} \, ,
\end{equation}
where $M^{exc}_{00}$ is the effective excitonic mass of state
\{00\}, {\em including} the electron-hole interaction, for $|\n n
|, |\n n'| \ll \hbar/r_H$. \cite{Lerner} The presence of the
delta function in (\ref{taza}) indicates that the scattering event
is kinematically elastic.

\section{Conclusions} \label{ultima}

We have presented explicit expressions for the residual
interaction between polarized excitons in strong magnetic
fields.  This potential would have important consequences on the
description of individual scattering events, as discussed, and the
collective excitations of the system.  Unfortunately, it is not
clear how one would perform experiments to directly measure the
many different scattering processes possible. Although we are
hopeful that some experiments might be designed in the future to
analyze these processes, we believe that the more direct probe
would be study of the various collective modes in this
interesting and unusual system.  Since the density-fluctuation
modes are now able to include rather complex internal
excitations, the resulting modes may indeed be quite unusual and
complicated.  One would use the potential derived here in an
approach similar to the case with no field, \cite{Dmitrii,Raul}
and results will be presented elsewhere.

We trust that these expressions would also be useful in the
description of other kinematic and thermodynamic properties of
the system.

 \acknowledgments This work was supported in part by
CONACYT-M\'{e}xico grant No.\ 983064, and US DOE No.\
DE--FG02--91ER45334.

\appendix

\section{Potential functions}

Because of the spurious short-range divergence introduced by the
$|\n x - \n x' |^{-3}$ dependence on the dipolar potential, we use
the physical cutoff parameter provided by the finite $z$-direction
polarization of the exciton $d$. \cite{Dmitrii}  Therefore, by
introducing the regularization factor as per $|\n x - \n x' |^{-3}
\rightarrow |\n x - \n x' |^{-3} (1-e^{-|\n x - \n x' |^2 /
d^2})$, we preserve the long-range dipolar interaction, while
allowing for the short-range Coulomb-like repulsion.
Correspondingly, we may write, with $\n S = \n x - \n x'$,
\begin{eqnarray}
  \phi (q) &=& \frac{1}{\kappa} \int \, e^{i \n q \cdot \n S}
  \frac{1}{S^3} (1-e^{-S^2 / d^2}) \, d^2 \n S \nonumber \\
 &=&  \frac{q}{2\pi \kappa} \left\{ -1 + \frac{qd\sqrt{\pi}}{4}
 e^{-q^2d^2/8} \left[ I_1 \left( \frac{q^2d^2}{8} \right) +
 \left( 1 + \frac{4}{q^2d^2} \right) I_0 \left( \frac{q^2d^2}{8}
 \right) \right]  \right\} \, ,
\end{eqnarray}
where $I_n$ is the Bessel function of imaginary argument.  Notice,
incidentally, that
\begin{equation}
 \phi (q \rightarrow 0) = \frac{1}{2 \kappa d \sqrt{\pi}} \, ,
\end{equation}
and is therefore ill-defined for $d=0$.  This function $\phi (q)$
accounts for the first term in the interaction in Eq.\
(\ref{term2}) or (\ref{final2}).

Similarly, the second dipolar term can be written as
\begin{eqnarray}
I ^{(2)} &=& \frac{1}{\kappa} \int e^{i\n q\cdot \n S} \frac{ (\n
p \cdot \n S) (\n p' \cdot \n S)}{S^5} (1 - e^{-S^4/d^4}) \, d^2
\n S \nonumber \\
 &=& \frac{1}{\kappa}\left(\n p \cdot \frac{1}{i}
\frac{\partial}{\partial \n q}\right) \left(\n p'\cdot \frac{1}{i}
\frac{\partial}{\partial \n q}\right) \int  e^{i\n q\cdot \n S}
\frac{1-e^{-\frac{S^4}{d^4}}}{S^5} \, d^2 \n S \nonumber \\
 &=& (\n p \cdot \n q) (\n p' \cdot \n q) \chi_1(q) -
 \chi _2(q) \, \n p \cdot \n p' \, , \label{intRR1}
\end{eqnarray}
where for $qd \gg 1$
\begin{equation}
\chi _1(q)= \frac{3}{\kappa} \sum_{k=0}^{\infty
}{\frac{2\pi(-256)^k (4k+1)(4k+3) \Gamma ^2(k+\frac{1}{4})\Gamma
^2(k+\frac{3}{4})} {\Gamma ^2(\frac{1}{4})\Gamma
^2(\frac{3}{4})(k+1)! d^{4k+4}q^{4k+5}}} \label{chi1}
\end{equation}
and
\begin{equation}
\chi _2(q)= \frac{3}{\kappa} \sum_{k=0}^{\infty
}{\frac{2\pi(-256)^k (4k+1) \Gamma ^2(k+\frac{1}{4})\Gamma
^2(k+\frac{3}{4})} {\Gamma ^2(\frac{1}{4})\Gamma
^2(\frac{3}{4})(k+1)! d^{4k+4}q^{4k+3}}} \label{chi2}
\end{equation}
On the other hand, for $qd \ll 1$, Eq.\ (\ref{intRR1}) behaves as,
\begin{equation}
I^{(2)} = -\frac{3\pi\Gamma(3/4)}{\kappa d}\n p\cdot\n p' \, .
\label{qpeq}
\end{equation}

\section{Dipole matrix elements}

Notice that the dipole moment matrix elements obey the symmetry
relation
\begin{equation}
{\n p}_{nm\n P,n'm'\n P'}={\n p}^{*}_{n'm'\n P',nm\n P} \, .
\end{equation}

Some special cases of overlap and dipole moment matrix elements
follow.  For the diagonal elements,
\begin{equation}
M_{00\n P',00\n P} = \exp \left[^{i\frac{\gamma}{4}\n q\cdot (\n
r_{\n P}+\n r_{\n P'}) - \frac{u^2}{8r_H^2} -
\frac{\gamma^2r_H^2q^2}{8}} \right] \, ,
\end{equation}
 where $\n u = \n r _{\n P} - \n r _{\n P'}$, and
 \begin{eqnarray}
{\n p}_{00\n P',00\n P}=\frac{e}{2}\left (\n r_{\n P}+\n r_{\n P'}
+ir_H^2\gamma\n q\right )\, e^{i\frac{\gamma}{4}\n q\cdot (\n
r_{\n P}+\n r_{\n P'})-(\gamma^2+1)\frac{r_H^2q^2}{8}}\nonumber\\
=\frac{e}{2}\left (\n r_{\n P}+\n r_{\n P'}+ir_H^2\gamma\n q\right
) \tilde{M}_{00\n P',00\n P}
\end{eqnarray}

The simplest/lowest off-diagonal elements are,
\begin{equation}
M_{01\n P',01\n P}=\left
(1-\frac{\gamma^2r_H^2q^2}{8}-\frac{u^2}{8r_H^2}+\frac{\gamma}{4}\,
\n u \times\n q \cdot \hat \n z\right )M_{00\n P',00\n P} \, ,
\end{equation}
 and
\begin{eqnarray}
{\n p}_{01\n P',01\n P}=\left [1-r_H^2q^2\left
(\frac{\gamma^2}{8}+\frac{\gamma}{4}+1\right )\right ] {\n
p}_{00\n P',00\n P}+i(\gamma+1)\frac{er_H^2\n q}{2}
\tilde{M}_{00\n P',00\n P} \, .
\end{eqnarray}

Similarly,
\begin{eqnarray}
M_{0-1\n P',0-1\n P} &=& \left
(1-\frac{\gamma^2r_H^2q^2}{8}-\frac{u^2}{8r_H^2}-\frac{\gamma}{4}\,
\n u \times\n q \cdot \hat \n z\right )M_{00\n P',00\n P} \, ,
\nonumber \\
 {\n p}_{0-1\n P',0-1\n P} &=& \left [1-r_H^2q^2\left
(\frac{\gamma^2}{8}+\frac{\gamma}{4}+\frac{1}{8}\right )\right ]
{\n p}_{00\n P',00\n P}+i(\gamma-1)\frac{er_H^2\n q}{2}
\tilde{M}_{00\n P',00\n P} \, ; \\ \nonumber \vspace{2em} \\
 M_{00\n P',01\n P} &=&
\frac{1}{\sqrt{2}}\left (i\frac{\gamma r_Hqe^{-i\varphi _q}}{2}-
\frac{u e^{-i\varphi _u}}{2r_H}\right ) M_{00\n P',00\n P} \, ,
\nonumber \\
 {\n p}_{00\n P',01\n P} &=& i\frac{\gamma+1}{2\sqrt{2}}r_{H}
[\n q\cdot (\hat\n x-i\hat\n y)]{\n p}_{00\n P',00\n
P}+\frac{er_H}{\sqrt{2}}(\hat\n x-i\hat\n y) \tilde{M}_{00\n
P',00\n P} \, ; \\  \vspace{2em} \nonumber \\
 M_{00\n P',0-1\n P} &=& \frac{1}{\sqrt{2}}\left (i\frac{\gamma
r_Hqe^{i\varphi _q}}{2}- \frac{u e^{i\varphi _u}}{2r_H}\right )
M_{00\n P',00\n P} \, , \nonumber \\
 {\n p}_{00\n P',0-1\n P} &=& i\frac{\gamma-1}{2\sqrt{2}}r_{H}[\n
q\cdot (\hat\n x+i\hat\n y)]{\n p}_{00\n P',00\n
P}+\frac{er_H}{\sqrt{2}}(\hat\n x+i\hat\n y) \tilde{M}_{00\n
P',00\n P}  \, ; \\  \vspace{2em} \nonumber \\
 M_{01\n P',0-1\n P} &=& -\left (\frac{\gamma^2r_H^2q^2e^{i2\varphi
_q}}{8} +\frac{u^2e^{i2\varphi _u}}{8r^2_H}\right ) M_{00\n
P',00\n P} \, ,  \nonumber \\
 {\n p}_{01\n P',0-1\n P} &=& -(\gamma^2+1)
 \frac{r^2_{H}q^2e^{i2\varphi_q}}{8} {\n p}_{00\n
P',00\n P}+i \frac{er^2_Hqe^{i\varphi_q}}{\sqrt{2}}(\hat\n
x+i\hat\n y) \tilde{M}_{00\n P',00\n P} \, .
\end{eqnarray}


\end{document}